\documentclass{article}
\usepackage{spconf,amsmath,graphicx,hyperref}
\usepackage{amsmath,amssymb,amsfonts}
\usepackage{graphicx}
\usepackage{textcomp}
\usepackage{booktabs}
\usepackage{threeparttable}
\usepackage{bm}
\usepackage{algorithm}
\usepackage{algpseudocode}

\usepackage{multicol,multirow}
\def\D{{\cal D}}

\def\L{{\cal L}}
\def\H{{\cal H}}

\usepackage{enumitem}

\title{Inverse-Hessian Regularization for Continual Learning in ASR}
\name{Steven Vander Eeckt \qquad Hugo Van hamme \thanks{Research supported by Research Foundation Flanders (FWO) under grant S004923N of the SBO programme.}}
\address{Department Electrical Engineering ESAT-PSI \\ KU Leuven, Leuven, Belgium \\
\texttt{\{steven.vandereeckt, hugo.vanhamme\}@esat.kuleuven.be}}
\begin{document}
\ninept
\maketitle
\begin{abstract}
Catastrophic forgetting remains a major challenge for continual learning (CL) in automatic speech recognition (ASR), where models must adapt to new domains without losing performance on previously learned conditions. Several CL methods have been proposed for ASR, and, recently, weight averaging—where models are averaged in a merging step after fine-tuning—has proven effective as a simple memory-free strategy. However, it is heuristic in nature and ignores the underlying loss landscapes of the tasks, hindering adaptability. In this work, we propose Inverse Hessian Regularization (IHR), a memory-free approach for CL in ASR that incorporates curvature information into the merging step. After fine-tuning on a new task, the adaptation is adjusted through a Kronecker-factored inverse Hessian approximation of the previous task, ensuring that the model moves primarily in directions less harmful to past performance, while keeping the method lightweight. We evaluate IHR on two CL benchmarks and show that it significantly outperforms state-of-the-art baselines, reducing forgetting while improving adaptability. Ablation studies and analyses further confirm its effectiveness.
\end{abstract}
\begin{keywords}
automatic speech recognition, continual learning, catastrophic forgetting, kronecker-factored approximation
\end{keywords}
\section{Introduction}
\label{sec:intro}

Automatic speech recognition (ASR) systems are widely deployed in everyday applications, from voice assistants to transcription services. 
To be accurate and inclusive, they must adapt to new speakers, accents, domains, or recording conditions. 
Yet such adaptation often causes \emph{catastrophic forgetting}~\cite{catastrophicforgetting}, where performance on previously learned conditions degrades. 
This severely limits the development of ASR models that can scale to diverse and evolving user populations. 
Continual learning (CL) addresses this challenge by enabling models to continually extend their knowledge.

CL for ASR has received increasing attention in recent years. 
\textit{Architectural methods} expand model capacity for new tasks~\cite{eeckt_adapters,sustek22_interspeech,disentangled}, 
while \textit{rehearsal-based methods} store small sets of past data in a memory for replay ~\cite{lifelongasr,eeckt2021continual}, which can be impractical due to storage or privacy constraints. 
\textit{Regularization-based methods} are memory-free and keep the architecture fixed; examples include updating only the encoder~\cite{updating_only}, adapting subsets of encoder layers~\cite{wang23d_interspeech}, or weight averaging~\cite{weight_averaging}. 
The latter—where the model is fine-tuned on a new task and then merged with the pre-adaptation model— has proven particularly successful, rivaling rehearsal-based methods while remaining simple and memory-free. However, weight averaging has two main limitations: (1) as the number of tasks grows, the weight of the most recent task shrinks toward zero, hindering adaptation; (2) the merging step is heuristic and ignores the loss landscapes of the tasks.

In this paper, we propose a new memory-free approach that incorporates loss landscape information into the merging step. 
Rather than naively averaging the pre- and post-adaptation models, we estimate the inverse Hessian of the previous tasks and use it to \textit{adjust} the adaptation update towards directions that are less sensitive for past tasks. 
This ensures that the model continues to learn the new task while remaining within low-loss regions of the old tasks. 
We approximate the Hessian separately per layer using Kronecker-factored approximations~\cite{kf,kronecker}.
Our contributions are threefold: (i) we introduce \emph{Inverse Hessian Regularization (IHR)}, a memory-free continual learning method for ASR that leverages curvature information in the merging step; 
(ii) we keep IHR lightweight and practical, as the inverse Hessian-vector multiplication is applied only once after fine-tuning; 
(iii) through two domain adaptation benchmarks, we demonstrate that IHR consistently outperforms strong baselines; while ablation and analysis confirm the effect of each component.

\section{Continual Learning for ASR}
\subsection{ASR Model}
We consider an encoder--decoder ASR model. 
Given an utterance $\bm{X} \in \mathbb{R}^{l \times d_i}$ of $l$ acoustic frames of dimension $d_i$, the model predicts a sequence $\hat{\bm{y}}$ of $\tilde{w}$ word pieces. 
Parameters are denoted by $\bm{\theta} \in \mathbb{R}^N$. 
The model is trained (and predicts) in a hybrid fashion \cite{hybrid_ctctransformer}, combining a CTC loss and a decoder cross-entropy loss with weights $c$ and $1-c$, resp. The loss is denoted by $\L(\bm X, \bm y; \bm \theta)$,
with $\bm y$ the ground truth sequence of $w$ tokens corresponding to $\bm X$.

\subsection{Continual Learning}
\label{subsec:problem}
\begin{figure}
    \centering
    \includegraphics[width=0.45\linewidth]{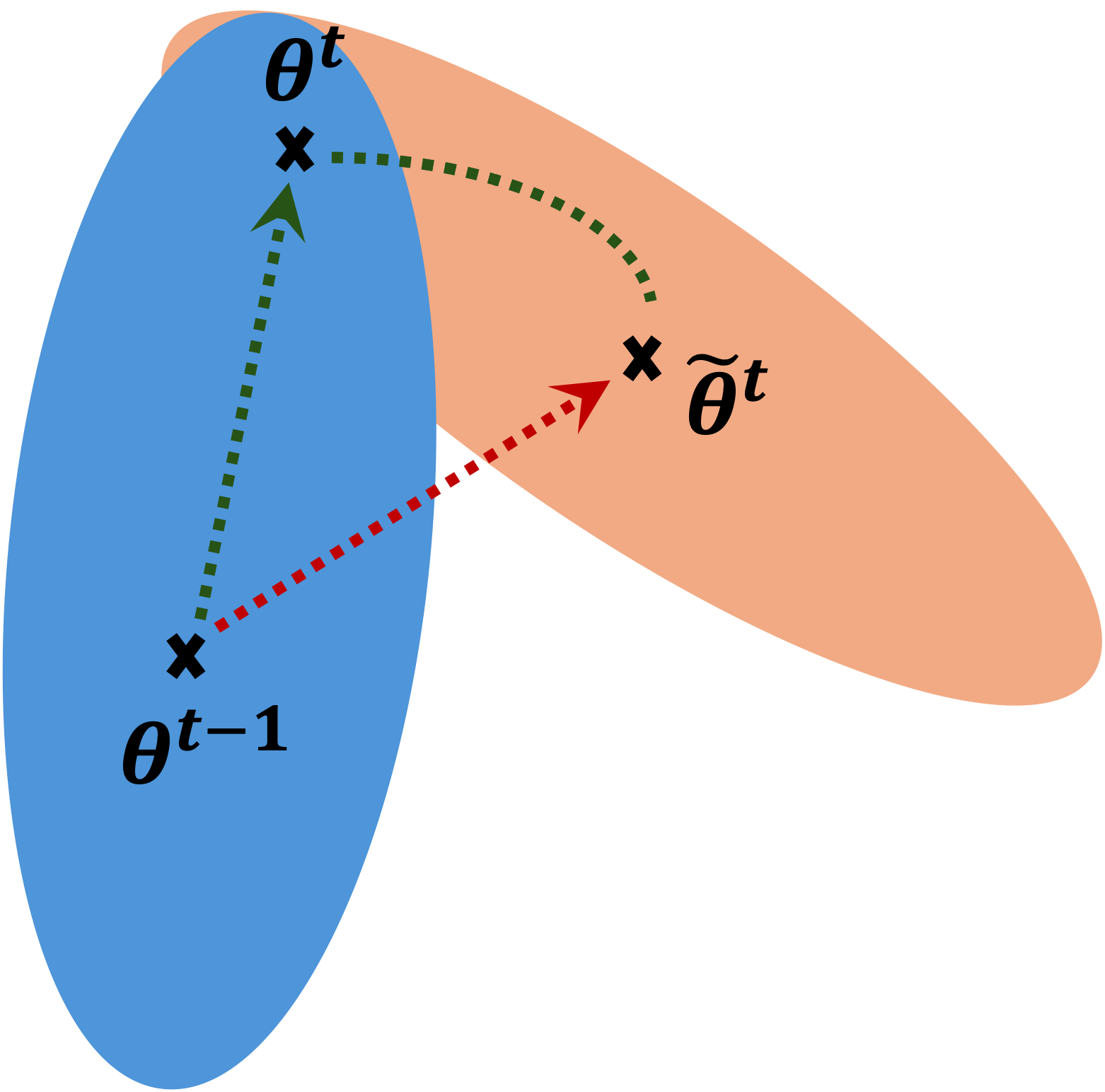}
\caption{Illustration of catastrophic forgetting. 
Fine-tuning moves the model from $\bm \theta^{t-1}$ to $\tilde{\bm \theta}^t$, 
entering the low-loss region of the new task $t$ (orange) but leaving that of tasks $1,\dots,t{-}1$ (blue). 
By multiplying the adaptation $\tilde{\bm \theta}^t - \bm \theta^{t-1}$ with the inverse Hessian,  resulting parameters $\bm \theta^{t}$ remain within the intersection of both low-loss regions.}
    \label{fig:loss_landscape}
\end{figure}

In continual learning (CL) for ASR, a model must learn a sequence of tasks $1,\dots,T$, which may reflect shifts in e.g. accent, dialect, language, speaker or recording conditions. For each task $t$, training data $\mathcal{D}_t$ is available, while past data $\mathcal{D}_{1:(t-1)}$ cannot be revisited once task $t$ starts. 
Given parameters $\bm{\theta}^{t-1}$ after task $t{-}1$, the goal is to obtain $\bm{\theta}^t$ that (i) preserves performance on previous tasks to minimize forgetting, and (ii) adapts effectively to the new task. 
Ultimately, after task $t$, the model should perform well on all tasks $1,\dots,t$. 
Naïve fine-tuning fails because it causes catastrophic forgetting; CL methods therefore modify optimization or training, adjust the architecture, or add a memory buffer with past samples.

%\subsection{Continual Learning}
%\label{subsec:problem}
%In CL for ASR, a model must learn a sequence of tasks $1,\dots,T$, which differ in domain (accent, speakers, etc.). 
%For each task $t$, training data $\mathcal{D}_t$ consisting of pairs $(\bm X, \bm y)$ is available, while past data $\mathcal{D}_{1:(t-1)}$ cannot be revisited once task $t$ starts. 

%Given parameters $\bm{\theta}^{t-1}$ after task $t{-}1$, the goal is to obtain $\bm{\theta}^t$ such that two criteria are balanced~\cite{biesialska-etal-2020-continual}:  
%\begin{enumerate}[label=(\roman*),leftmargin=*]
%    \item \textit{Preserve past knowledge:} performance on previous tasks should not degrade, i.e. catastrophic forgetting is minimized. 
%    \item \textit{Learn the new task:} the model should adapt effectively to $\mathcal{D}_t$, e.g. 
%    \begin{equation}
%        \bm{\theta}^t \approx \arg\min_{\bm{\theta}} 
%        \sum_{(\bm{X},\bm{y})\in\mathcal{D}^\text{train}_t}
%        \mathcal{L}(\bm{X},\bm{y};\bm{\theta}).
%    \end{equation}
%\end{enumerate}
%Satisfying these criteria requires that after task $t$ the model with parameters $\bm \theta^{t}$ performs well on all tasks $1,\dots,t$. To achieve this, only the parameters $\bm{\theta}^{t-1}$ and the new data $\mathcal{D}_t$ are available. 
%Naïvely fine-tuning $\bm{\theta}^{t-1}$ on $\mathcal{D}^\text{train}_t$ causes catastrophic forgetting; To overcome this, CL methods typically modify the optimization procedure, adjust the architecture, or incorporate a memory buffer containing samples from previous tasks.

\section{Inverse Hessian Regularization}
After fine-tuning the model on task $t$ using its training data $\D_t$, starting from parameters $\bm{\theta}^{t-1}$, we obtain updated parameters $\bm{\tilde{\theta}}^t$. 
These parameters may lie outside the low-loss region of tasks $1,\dots,t{-}1$, leading to forgetting (illustrated in Fig.~\ref{fig:loss_landscape}). 
To mitigate this, we introduce a merging step that adjusts the adaptation update through an inverse Hessian approximation of the previous task(s), resulting in parameters $\bm{\theta}^t$ that better balance old and new tasks.

\subsection{Merging Step: Inverse Hessian Regularization}
\label{subsec:reasoning}

Ideally, while adapting to task $t$, the model should remain within the low-loss regions of the earlier tasks. This would require updates to occur mainly along directions to which tasks $1,\dots,t{-}1$ are insensitive, i.e., directions that leave their performance unchanged. Such directions can be identified from the inverse Hessian of the loss on tasks $1,\dots,t{-}1$ around $\bm \theta^{t-1}$: its dominant eigenvectors, corresponding to the largest eigenvalues, represent directions in which the loss landscape of previous tasks is relatively flat.

Instead of incorporating the inverse Hessian directly as a regularizer during training \cite{ncl}, it can also be applied in a post-processing or merging step, keeping the method lightweight as the regularization is performed only once after fine-tuning. Specifically, the update $\Delta \bm \theta^t = \tilde{\bm \theta}^t - \bm \theta^{t-1}$ can be multiplied with the inverse Hessian, so that the final parameters become
\begin{equation}
\bm \theta^{t} = \bm \theta^{t-1} + \bm \H_{1:t-1}^{t-1} \Delta \bm \theta^{t},
\label{eq:inverse_hessian_full_model}
\end{equation}
where $\bm \H_{1:t-1}^{t-1}$ denotes the inverse Hessian of the loss on tasks $1,\dots,t{-}1$ evaluated at $\bm \theta^{t-1}$. The inverse Hessian rescales the parameter update $\Delta\bm \theta^t$ and is therefore used as part of an update rule acting directly in parameter space.

There are two problems with this approach, however, as the computation of $\bm \H_{1:t-1}$ is infeasible for two reasons:

\textbf{First}, computing the full Hessian $\bm \H_{1:t-1}^{t-1} \in \mathbb{R}^{N \times N}$ is infeasible except for unrealistically small models with few parameters $N$. However, the Hessian can be approximated. In Elastic Weight Consolidation \cite{ewc}, this is done by considering the diagonal of the Fisher information matrix. This, however, makes the unrealistic assumption that model parameters are independent. To improve on this, \cite{csqn} considers Quasi-Newton methods, capturing parameter interactions in a low-rank approximation, while \cite{kf} uses Kronecker-factored approximations \cite{kronecker}, which assume independence across layers but not within them. \textit{We adopt the latter approach and compute Eq.~\ref{eq:inverse_hessian_full_model} at the layer rather than the full-model level}.

%\textbf{Second}, even if computing the full Hessian $\bm \H_{1:t-1}^{t-1}$ were feasible, it could not be done in practice as access to $\D_{1:t-2}$ is lost when $\bm \theta^{t-1}$ is obtained. At that point, only the Hessian $\bm \H^{t-1}_{t-1}$ can be computed. Therefore, \textit{we approximate $\bm \H_{1:t-1}^{t-1}$ in one of two ways}: (1) by summing per-task Hessians, $\bm \H_{1:t-1}^{t-1} \gets \sum_{i=1}^{t-1} \bm \H_i^i$, as in \cite{ewc_more, csqn}; or (2) by using only the most recent task, i.e., $\bm \H_{1:t-1}^{t-1} \gets \bm \H_{t-1}^{t-1}$.

\textbf{Second}, even if computing the full Hessian $\bm \H_{1:t-1}^{t-1}$ were feasible, it could not be done in practice as access to $\D_{1:t-2}$ is lost when $\bm \theta^{t-1}$ is obtained. At that point, only the Hessian $\bm \H^{t-1}_{t-1}$ can be computed. Therefore, \textit{we approximate $\bm \H_{1:t-1}^{t-1}$ by using only the most recent task}, i.e., $\bm \H_{1:t-1}^{t-1} \gets \bm \H_{t-1}^{t-1}$.

\subsection{Practical Layerwise Implementation}
Using Kronecker-factored block-diagonal approximations to estimate the Hessian, we apply Eq.~\ref{eq:inverse_hessian_full_model} at the layer level. Since these approximations are particularly effective for linear layers \cite{csqn}, we restrict the regularization to the linear layers, which account for the vast majority of parameters and are therefore assumed to be most critical for mitigating interference between old and new tasks.

Consider a linear weight matrix $\bm W \in \mathbb{R}^{d_o \times d_i}$ with bias $\bm b \in \mathbb{R}^{d_o}$, producing output $\bm W \bm x + \bm b$ for input $\bm x \in \mathbb{R}^{d_i}$. 
Let $\bm W_{t-1}$ from $\bm \theta^{t-1}$ and $\tilde{\bm W}_t$ from $\tilde{\bm \theta}^t$ denote its parameters before and after fine-tuning on task $t$, respectively. 
The latter can be written as $\tilde{\bm W}_t = \bm W_{t-1} + \Delta \bm W_t$, where $\Delta \bm W_t \in \mathbb{R}^{d_o \times d_i}$ is the task-specific update.

If we denote by $\bm H_{t-1}^{t-1}$ the inverse Hessian of $\L(\bm X, \bm y; \bm \theta^{t-1})$ 
with respect to the parameters $\bm W_{t-1}$ on data $(\bm X, \bm y) \in \D_{t-1}$, 
then, following Eq.~\ref{eq:inverse_hessian_full_model}, we adjust the update $\Delta \bm W_t$ using $\bm H_{t-1}^{t-1}$ as follows:
\begin{equation}
   \bm{W}_t = \bm{W}_{t-1} + \alpha \,\bm H_{t-1}^{t-1} \Delta \bm{W}_t
   \label{eq:inverse_hessian_practice}
\end{equation}
The scalar factor $\alpha$ rescales the adjusted update to have a comparable norm to the original update, and is defined as
\begin{equation}
    \alpha = \tau \frac{\|\Delta \bm W_t \|}{\|\bm H_{t-1}^{t-1} \Delta \bm{W}_t\|}
    \label{eq:alpha}
\end{equation}
where $\tau$ is a tunable hyper-parameter, providing an adjustable trade-off between stability and plasticity.

As explained in Section~\ref{subsec:reasoning}, we approximate $\bm H_{1:t-1}^{t-1}$, whose computation is infeasible, by $\bm H_{t-1}^{t-1}$, meaning that {only one inverse Hessian estimate (i.e. only for task $t-1$) needs to be stored}. 
For each linear layer $\bm W$, using Kronecker-factored approximations, the Hessian estimate requires the storage of a $d_o \times d_o$ and a $d_i \times d_i$ matrix. Since only the most recent task's inverse Hessian is retained, \textit{storage requirements remain constant as $t$ grows}, a key requirement in CL~\cite{biesialska-etal-2020-continual}.
Moreover, the inverse Hessian--vector multiplication from Eq.~\ref{eq:inverse_hessian_practice} is \textit{applied only once per task, during the merging step}. 
Consequently, \textit{the method remains both storage efficient and lightweight}.

\subsection{Handling Remaining Parameters}
\label{subsec:remaining}
While most parameters reside in the linear layers, others remain outside them (e.g., in convolutional layers, normalization layers, or biases). 
For such a parameter $\bm p$, let $\bm p_{t-1}$ denote its value in $\bm \theta^{t-1}$ and $\tilde{\bm p}_t = \bm p_{t-1} + \Delta \bm p_t$ its value in $\tilde{\bm \theta}^t$. 
We then define a scalar $\alpha_p$ such that the final parameter after task $t$ ($\bm p_t$ from $\bm \theta^t$) becomes $\bm p_t = \bm p_{t-1} + \alpha_p \Delta \bm p_t$. Following \cite{weight_averaging}, we consider $\alpha_p=1/t$.

\subsection{Overview}
Our method, named Inverse Hessian Regularization (IHR), consists two parts. First, the model is fine-tuned on the new task $t$ starting from parameters $\bm \theta^{t-1}$, yielding parameters $\tilde{\bm \theta}^t$. 
Next, we correct the task-specific update $\Delta\bm\theta^t = \tilde{\bm \theta}^t - \bm \theta^{t-1}$ through multiplication with an inverse Hessian approximation estimated at $\bm \theta^{t-1}$, so that the final parameters $\bm \theta^t$ move primarily in directions that are less sensitive for previous tasks. This step is performed using Kronecker-factored inverse Hessian estimates \cite{kronecker} for the linear layers, accounting for most parameters. 
For remaining parameters (e.g., convolution, normalization, bias), we apply a simple scalar weighting of their updates. 
Algorithm~\ref{alg:ihvp} summarizes the procedure.

\begin{algorithm}
\caption{Inverse Hessian Regularization (IHR) for CL}
\label{alg:ihvp}
\begin{algorithmic}[1]
\Require Previous model $\bm{\theta}^{t-1}$, data $\mathcal{D}_t$ for new task $t$, $\tau$ and $\alpha_p$
\State \textbf{Fine-Tuning:} Fine-tune $\bm{\theta}^{t-1}$ on $\mathcal{D}_t$ to obtain $\tilde{\bm{\theta}}^t$
\State \textbf{Merging Step - Inverse Hessian regularization: }
\For{each linear layer with weights $\bm W$}
    %\State \textit{\# $\bm W_{t-1}$ and $\tilde{\bm W}_t$ are $\bm W$ from $\bm \theta^{t-1}$ and $\tilde{\bm \theta}^t$, resp.}
    \State Compute update: $\Delta \bm W_t = \tilde{\bm W}_t - \bm W_{t-1}$
    \State Adjust update: $\bm W_t = \bm W_{t-1} + \alpha \,\bm H_{t-1}^{t-1}\,\Delta \bm W_t$ \Comment{Eq. \ref{eq:inverse_hessian_practice}}
\EndFor
\For{each remaining parameter $\bm p$ (e.g., conv, norm, bias)}
    %\State \textit{\# $\bm p_{t-1}$ and $\tilde{\bm p}_t$ are $\bm p$ from $\bm \theta^{t-1}$ and $\tilde{\bm \theta}^t$, resp.}
    %\State Compute update $\Delta \bm p_t=\tilde{\bm p}_{t} - \bm p_{t-1}$ 
    %\State Average: $\bm p_t = \bm p_{t-1} + \alpha_p \Delta \bm p_t$ \Comment{Eq. \ref{eq:rest_of_model}}
    \State Average: $\bm p_t = \bm p_{t-1} + \alpha_p (\tilde{\bm p}_{t} - \bm p_{t-1})$ \Comment{Sec. \ref{subsec:remaining}}
\EndFor
%\State Estimate layer-wise inverse Hessians $\bm H_{t}^{t}$ using Kronecker factors on $\mathcal{D}_t$ with the merged model parameters
\State Estimate inverse Hessians $\bm H_{t}^{t}$, layer-wise for each $\bm W$, using Kronecker factors on $\mathcal{D}_t$ with the merged model parameters
\State \Return Final model parameters $\bm{\theta}^t = \{\bm W_t, \bm p_t, \dots\}$
\end{algorithmic}
\end{algorithm}

\section{Experiments}
\begin{table*}
    \centering
    \begin{threeparttable}
    \caption{Results of the experiments. Tasks are learned from left to right, with WERs obtained after learning all tasks. Best WER per column (across memory-free CL methods) is in bold, second best underlined. $\dagger$ indicates that the method relies on storage of past data.}
    \begin{tabular}{l c@{\hspace{4pt}} c@{\hspace{4pt}} c@{\hspace{4pt}} c@{\hspace{4pt}} c@{\hspace{4pt}} c@{\hspace{4pt}} c@{\hspace{4pt}}  c@{\hspace{4pt}} c@{\hspace{4pt}} c@{\hspace{4pt}} c@{\hspace{4pt}} c@{\hspace{4pt}} c@{\hspace{4pt}} c@{\hspace{4pt}}}
    \toprule
    & \multicolumn{7}{c}{\textbf{\textit{Exp. 1}}} & \multicolumn{7}{c}{\textbf{\textit{Exp. 2}}} \\
    \cmidrule(lr){2-8} \cmidrule(lr){9-15}
    \multirow{2}{*}{\textbf{Method}} & \multicolumn{5}{c}{\textbf{WER$\downarrow$ per task}} & \multicolumn{2}{c}{\textbf{Average}} & \multicolumn{5}{c}{\textbf{WER$\downarrow$ per task}} & \multicolumn{2}{c}{\textbf{Average}}   \\
    \cmidrule(lr){2-6} \cmidrule(lr){7-8}  \cmidrule(lr){9-13} \cmidrule(lr){14-15}
      & \textbf{1--US} & \textbf{2--ENG} & \textbf{3--AUS} & \textbf{4--IND} & \textbf{5--SCO} & \textbf{WER}$\downarrow$  & \multicolumn{1}{c}{\textbf{BWT}$\uparrow$}  & \textbf{1--LIB} & \textbf{2--GB/M} & \textbf{3--US/U} & \textbf{4--IN/U} & \textbf{5--IN/M} & \textbf{WER}$\downarrow$  & \textbf{BWT}$\uparrow$ \\
    \toprule
      \multicolumn{1}{l}{Initial model} & 15.4 & 12.7 & 13.4 & 21.4 & 13.5 & 15.25 & --- & \phantom{1}6.4 & 19.4 & \phantom{1}7.9 & 18.8 & 32.9 & 17.08 & --- \\
    \multicolumn{1}{l}{Fine-Tuning} & 18.2 & 12.1 & 12.4 & 22.2 & {10.5} & 15.07 & -3.6 & 17.0 & 19.6 & 16.5 & \phantom{1}4.1 & \phantom{3}4.9 & 12.43 & -9.0 \\
    %\multicolumn{1}{l}{Sep. Model} & 15.4 & 10.2 & \phantom{1}8.9 & 16.0 & 10.5 & 12.17 & \phantom{1}0.0 & \phantom{1}6.4 & \phantom{1}4.7 & \phantom{1}5.9 & \phantom{1}4.3 & \phantom{1}4.9 & \phantom{1}5.23 & \phantom{1}0.0 \\
     ER$\dagger$ & 17.0 & 11.3 & 11.8 & 19.3 & 10.4 & 13.97 & -2.3 & \phantom{1}7.6 & \phantom{1}6.8 & \phantom{1}7.9 & \phantom{1}4.5 & \phantom{1}5.4 & \phantom{1}6.43 & -1.1 \\
    \midrule
    \multicolumn{1}{l}{FTA} & \underline{15.9} & \underline{10.9} & \textbf{10.7} & \underline{19.3} & 11.8 & 13.71 & -0.3 & \phantom{1}\textbf{7.5} & \phantom{1}\textbf{8.6} & \phantom{1}\textbf{8.1} & \phantom{1}8.5 & 12.2 & \phantom{1}8.97 & -0.1 \\
    \multicolumn{1}{l}{UOE} & 18.5 & 12.5 & 12.8 & 22.6 & \textbf{10.5} & 15.36 & -3.8 & 14.3 & 19.4 & 17.1 & \phantom{1}\textbf{4.4} & \phantom{1}\textbf{5.4} & 12.10 & -8.2 \\
    \multicolumn{1}{l}{CLRL-T} & 17.9 & 12.2 & 12.4 & 22.3 & 11.6 & 15.26 & -2.8 & 11.7 & 16.4 & 14.4 & \phantom{1}5.6 & \phantom{1}7.1 & 11.02 & -5.5 \\
    %EWC & \\
    %IMM & \\
    %\cmidrule(lr){1-15}
      %ER [$|\M|=20$] & 17.8 & 11.8 & 12.3 & 20.8 & 10.5  & 14.62\tnote{e,f} & -3.0 & \phantom{1}9.4 & 10.4 & \phantom{1}9.9 & \phantom{1}4.3 & \phantom{1}5.1 & \phantom{1}7.82\tnote{d} & -3.0 \\
      %ER [$|\M|=1t$] & 18.2 & 11.8 & 12.5 & 21.0 & 10.5 & 14.80\tnote{e} & -3.2 & 12.2 & 13.7 & 11.8 & \phantom{1}4.2 & \phantom{1}5.1 & \phantom{1}9.39\tnote{d} & -5.1 \\
    %\cmidrule(lr){1-15}
     %SVR & 15.9 & 10.4 & 10.4 & 18.3 & 10.9 & \textbf{13.18}\tnote{a,f} & -0.6 & \phantom{1}8.3 & \phantom{1}7.9 & \phantom{1}8.8 & \phantom{1}5.8 & \phantom{1}7.7 & \phantom{1}\underline{7.70}\tnote{b,f} & -1.1 \\
    %SVR [$|\M|=1t$] & 16.0 & 10.5 & 10.5 & 18.6 & 11.0 & \underline{13.33}\tnote{a,f} & -0.7 & \phantom{1}8.4 & \phantom{1}8.2 & \phantom{1}8.7 & \phantom{1}5.5 & \phantom{1}7.3 & \phantom{1}\textbf{7.62}\tnote{c,f}  & -1.3 \\
    \midrule 
    IHR & \textbf{15.0} & \textbf{10.4} & \textbf{10.7} & \textbf{19.0} & \underline{11.5} & \textbf{13.32}\tnote{a} & -0.1 & \phantom{1}\underline{8.1} & \phantom{1}\underline{8.8} & \phantom{1}\underline{8.9} & \phantom{1}\underline{5.0} & \phantom{1}\underline{6.2} & \phantom{1}\textbf{7.40}\tnote{a} & -1.4  \\
    \bottomrule
    \end{tabular}
    \begin{tablenotes}
    \footnotesize
    \item[a] Significantly outperforms all relevant baselines (i.e. those not relying on past data) with significance level $0.001$.
    \end{tablenotes}
    \label{tab:full_results}
    \end{threeparttable}
\end{table*}

Experiments are done in ESPnet2 \cite{watanabe2018espnet}. More information, including code and detailed results, can be found in our Github repository \footnote{{https://github.com/StevenVdEeckt/inverse-hessian-regularization}}.

\noindent \textbf{Data.} We consider two CL benchmarks:
(\textit{Exp.~1}) Following~\cite{weight_averaging}, we use Common Voice (CV)~\cite{commonvoice} English data set, divided into five accents: United States (US), England (ENG), Australia (AUS), India (IND), and Scotland (SCO). Tasks are presented sequentially in this order. 
(\textit{Exp.~2}) Following \cite{vandereeckt24_interspeech}, we start from a model trained on LibriSpeech-360h (LIB)~\cite{librispeech},  and adapt it to four Libri-Adapt ~\cite{libri_adapt} tasks, involving dual domain shifts in microphone type (USB [U], Matrix [M]) and accent (US, India [IN], Britain [GB]). 

\noindent \textbf{Model.} The model consists of 12 Conformer encoder blocks~\cite{conformers} and 6 Transformer decoder blocks~\cite{transformer}. Each block has 4 attention heads of dimension 256 and a feed-forward dimension 2048. SentencePiece~\cite{sentencepiece} is employed to construct a vocabulary of $5000$ subword tokens, generated on the first task. During training and decoding, the CTC weight is set to $c=0.3$. 
The model has 46.7M parameters, of which 90.7\% lie in the weight matrices of linear layers. Optimization is performed with Adam~\cite{adam}, re-initialized before each new task. 
The initial task is trained for 80 epochs; subsequent tasks ($2,\dots,T$) for 10 epochs each, with a learning rate reduced by a factor of 10 compared to the initial task, following~\cite{eeckt2021continual}.

\noindent\textbf{Baselines.} 
We compare against: (1) \textit{Initial model}, i.e., with parameters $\bm{\theta}^1$ before adaptation; 
(2) \textit{Fine-Tuning}, naive adaptation with forgetting (worst case); 
%(3) \textit{Separate Model}, a task oracle with one model per task (best case); 
(3) \textit{Experience Replay (ER)}~\cite{er}, which jointly trains on current data and stored utterances in a memory. 
Although ER violates the memory-free constraint, it serves as a reference for methods without access to past data; 
(4) \textit{Fine-Tuning with Averaging (FTA)}~\cite{weight_averaging}, averaging $\bm \theta^{t-1}$ and $\tilde{\bm \theta}^t$ with $\eta=1/t$; 
(5) \textit{Updating Only Encoder}~\cite{updating_only}, freezing decoder and norm layers; 
(6) \textit{CLRL-T}~\cite{wang23d_interspeech}, updating $K$ random encoder layers per epoch. We take $K=1$, best setting in~\cite{wang23d_interspeech}, and train for 50 epochs. 
Hyper-parameters are tuned using the validation sets on the first adaptation. For our method, IHR, this results in $\tau=1$ for Exp. 1 and $\tau=3$ for Exp. 2. For ER, the memory size is 2000 utterances for Exp. 1 and 200 for Exp. 2.

%\noindent\textbf{Baselines}. We compare our method to the following baselines: (1) \textit{Initial model}: the model prior to adaptation with parameters $\bm{\theta}^1$; (2)  \textit{Fine-Tuning}: naively adapts the model to the subsequent tasks, represents the worst-case scenario; %(3) \textit{Seperate Model (Sep. Model)}: keeps a separate model for each task combined with a task oracle, avoiding task interference and forgetting -- it represents a best-case scenario; 
%(3) \textit{Experience Replay (ER)} \cite{er}: trains jointly on the current mini-batch from the new task and on a mini-batch sampled from the memory; (5) \textit{Fine-Tuning with Averaging (FTA)} \cite{weight_averaging}: Averages the model before $\bm \theta^{t-1}$ and after adaptation $\tilde{\bm \theta}^{t}$ with weights $(1-\eta)$ and $\eta$, resp., with $\eta=1/t$; (6) Updating Only Encoder \cite{updating_only}: freezes the decoder and the norm layers for adaptation; (7) \textit{CL by Random Layer-wise Tuning (CLRL-T)} \cite{wang23d_interspeech}: proposes to only update $K$ randomly selected encoder layers each epoch -- we set $K=1$ (best setting in \cite{wang23d_interspeech} and increase epochs to 50). To determine the hyper-parameters of ER and our method, we use the validation sets of the first adaptation. This results in $\tau=1$ for CV and $\tau=3$ for LIB-APT experiments. For ER, the memory size is set at 2000 for CV and 200 for LIB-APT. 

\noindent \textbf{Metrics.} 
We report WER (in \%) per task, evaluated on the final model trained on all tasks. 
Overall performance is assessed through 
$\text{Average WER} = \tfrac{1}{T}\sum_{k=1}^{T} R_{T,k}$, 
where $R_{i,j}$ denotes the WER on task $j$ after training on the first $i$ tasks. 
We also report Backward Transfer (BWT) \cite{eeckt2021continual}, defined as 
$\text{BWT} = \tfrac{1}{T-1}\sum_{k=1}^{T-1} (R_{k,k} - R_{T,k})$, 
with negative values indicating forgetting.

\noindent \textbf{Significance testing}. We test significance on Avg. WER using Wilcoxon signed-rank test on per-utterance errors \cite{Strik2000ComparingTR}.

\section{Results}

\subsection{Experiment 1}
As shown by Table \ref{tab:full_results}, our method (IHR) significantly outperforms all baselines, being able to learn with close to zero forgetting (as shown by its -0.1 BWT). In addition, we observe the following:

\textbf{First}, compared to FTA, the strongest memory-free baseline, IHR achieves a clear performance gain.  The improvement is partly due to slightly reduced forgetting, but more importantly due to substantially better adaptability to new tasks. 
This can be seen from the last task, 5--SCO, where IHR achieves a reduction of 0.3 WER points compared to FTA. 
The difference is also evident in the Average WER: the reduction in BWT from $-0.3$ (FTA) to $-0.1$ (IHR) explains only a limited part of the 0.4 Avg. WER point improvement; the remainder stems from improved learning of the new tasks. 

\textbf{Second}, IHR also surpasses ER, despite ER having access to stored utterances from past tasks. IHR reduces the latter's forgetting by more than $90\%$, without violating any privacy constraints.

\textbf{Third}, looking at tasks individually, IHR attains the best result on four out of the five accents. 
On the remaining task, 5--SCO, it ranks second behind UOE, which, however, suffers from even stronger forgetting than Fine-Tuning. 

\textbf{Fourth}, notably, on the first task 1–US, IHR achieves a WER of 15.0, lower than the initial 15.4 from the Initial model $\bm \theta^1$, indicating positive BWT \cite{biesialska-etal-2020-continual}: information from later tasks is exploited to enhance earlier ones. Thanks to this positive BWT on 1--US, IHR, with its limited forgetting and high adaptability, is the only method improving performance on \emph{all five} tasks compared to \textit{Initial model}. %$\bm \theta^1$. 

\subsection{Experiment 2}
Table \ref{tab:full_results} shows that, similar to Exp.~1, IHR achieves the best overall performance with limited forgetting. Several trends can be observed. 

\textbf{First}, forgetting is generally higher in this experiment, due to the additional microphone domain shift, as indicated by more negative BWT across methods. This is most pronounced for Fine-Tuning ($-9.0$ vs.\ $-3.6$ in Exp.~1). Our method also forgets more than in Exp.~1, but the effect remains limited compared to the baselines. 

\textbf{Second}, while its forgetting is slightly higher than for FTA, IHR's overall performance is much stronger, reducing FTA's average WER by more than $17\%$. In this more challenging experiment, the low adaptability of FTA becomes problematic, whereas our method maintains a good balance between learning new tasks and retaining old ones. UOE shows the opposite behavior, adapting well but failing to prevent forgetting. The ability of our method to find, unlike FTA, UOE or CLRL-T, a compromise, is also reflected in the per-task results: FTA performs best on the first three tasks but poorly on the last two, UOE excels on the last two but catastrophically forgets the first three, while our method is consistently second-best on all five tasks, always within 1.0 WER of the best score. 

\textbf{Third}, although our method outperforms all memory-free baselines, ER remains strongest in Exp. 2. Since utterances in Exp. 2 are much longer than in Exp. 1, fewer examples suffice for ER to be highly effective, which explains its reduced forgetting compared to Exp.~1. Of all memory-free methods, however, our method comes closest to ER's performance with only slightly higher forgetting. 

\subsection{Analysis: Effect of the Inverse Hessian Regularization}
\begin{figure}
    \centering
    \includegraphics[width=0.92\linewidth]{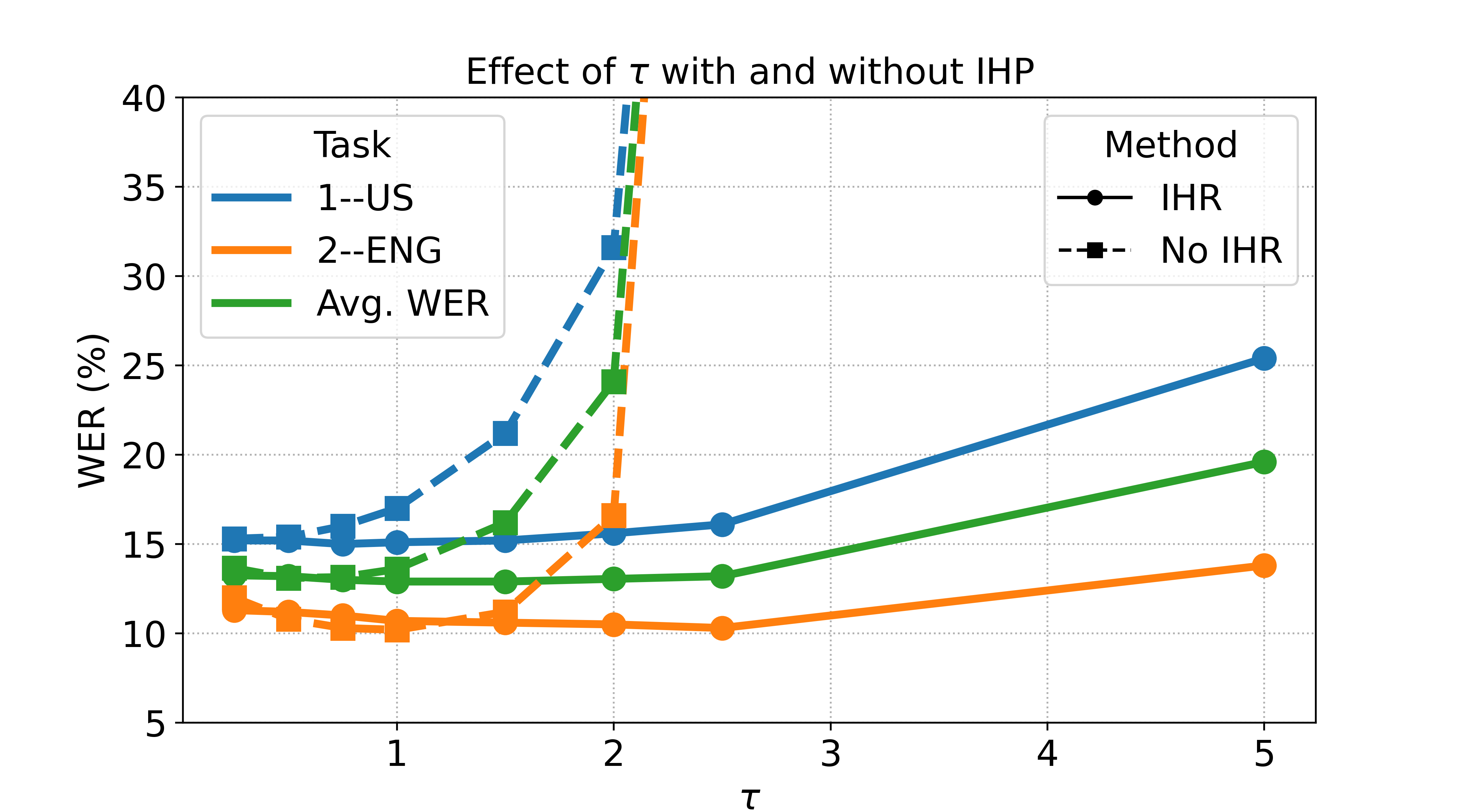}
\caption{ WER as a function of the hyper-parameter $\tau$. 
Colors indicate the reported task (1--US, 2--ENG, or Average WER), 
while line style and marker indicate whether inverse Hessian Regularization (IHR) is applied (solid, circles) or not (dashed, squares).}
    \label{fig:analysis}
\end{figure}

To illustrate that inverse Hessian regularization steers updates into directions that are less harmful for previous tasks, 
Figure~\ref{fig:analysis} plots the WER as a function of $\tau$ (applied through Eq.~\ref{eq:inverse_hessian_practice} and Eq.~\ref{eq:alpha}) 
for the first adaptation step (1--US $\rightarrow$ 2--ENG) in Exp. 1. 
For comparison, the WER obtained by moving directly along $\Delta \bm W_t$ with step size $\tau$, i.e., without applying the inverse Hessian (\textit{No IHR}), is also shown.
As illustrated, applying the inverse Hessian regularization allows $\tau$ to be increased while maintaining good performance on both tasks. 
WER only begins to rise after $\tau\!=\!2.5$, and remains stable even at $\tau\!=\!5.0$. 
In contrast, without regularization the WER deteriorates rapidly from $\tau\!=\!2.0$ onward, most severely on the old task.

\subsection{Ablation Study}
\label{subsec:ablation}
Table \ref{tab:ablation} provides an ablation study of our method. Comparing the second to the third row, it can be seen that using $\bm H_{t-1}^{t-1}$, rather than $\sum_{i=1}^{t-1} \bm H_{i}^i$, which \cite{ewc,kf, csqn} use to approximate $\bm H_{1:t-1}^{t-1}$, does not underperform the latter. Meanwhile, the former has the advantage of being more storage efficient. The third and fourth row, on the other hand, illustrate the effect of the remaining $9.3\%$ of the parameters: averaging them with $\alpha_p=1/t$, as done in FTA \cite{weight_averaging}, further reduces the forgetting, significantly improving the overall performance.

\begin{table}
    \centering
    \begin{threeparttable}
    \caption{Ablation study of our method. "+" indicates modifications are made compared to the line directly above.}
    \begin{tabular}{l l r}
    \toprule
    \multirow{2}{*}{\textbf{Method}} & \multicolumn{2}{c}{\textbf{Average}} \\
    \cmidrule(lr){2-3}
     & \textbf{WER}$\downarrow$ & \textbf{BWT}$\uparrow$  \\
    \toprule
   Fine-Tuning & 15.07 & -3.6 \\
    + IHR using $\sum_{i=1}^{t-1} \bm H_{i}^i$ and $\alpha_p=0.50$  & 13.54\tnote{a} & -0.6 \\
    + use $\bm H_{t-1}^{t-1}$ instead of $\sum_{i=1}^{t-1} \bm H_{i}^i$ & 13.53 & -0.6 \\
    + use $\alpha_p=1/t$ instead of $\alpha_p=0.50$ & 13.32\tnote{b} & -0.1 \\ 
     \bottomrule    
    \end{tabular}
    \begin{tablenotes}
    \footnotesize
    \item[a, b] Significant improvement w.r.t. the method above, with significance levels $0.001$ (a) and $0.01$ (b).
    \end{tablenotes}
    \label{tab:ablation}
    \end{threeparttable}
\end{table}

\section{Discussion}
In ASR, tasks are often highly similar, and many parameters are simultaneously important for both past and new domains. 
As shown in~\cite{eeckt2021continual}, this makes \textit{prior-focused methods}~\cite{defy,ewc,csqn,mas}, which rely on an $L_2$ penalty weighted by a Hessian approximation, ill-suited: the model faces no good compromise, either updating these parameters and forgetting, or freezing them and failing to learn. 
Weight averaging~\cite{weight_averaging} avoids catastrophic forgetting by interpolating between old and new models, but applies compromise uniformly across all parameters, even where it is unnecessary. 
IHR differs fundamentally: by adjusting the update through the inverse Hessian, we hypothesize that it gently rescales updates in sensitive directions rather than blocking them, while preserving those needed to learn new tasks; a balance confirmed by our empirical results.
%At the same time, IHR remains lightweight: only one inverse Hessian estimate is kept, and the multiplication is applied once per task, combining effectiveness with efficiency. 
%The empirical results confirm this balance: IHR improves adaptability while maintaining minimal forgetting in both benchmarks. 
%Although our work focuses on ASR, the same principle could benefit other continual learning problems where privacy, storage, and adaptability must be jointly addressed.

%\section{Discussion}
%In ASR, tasks are often highly similar, and many parameters are simultaneously important for both past and new domains. 
%As shown in~\cite{eeckt2021continual}, this makes \textit{prior-focused methods}~\cite{defy,ewc,csqn}, which rely on an $L_2$ penalty weighted by a Hessian approximation, ill-suited: the model faces no good compromise, either updating these parameters and forgetting, or freezing them and failing to learn. 
%Weight averaging~\cite{weight_averaging} avoids catastrophic forgetting by interpolating between old and new models, but applies compromise uniformly across all parameters, even where it is unnecessary. 
%IHR differs fundamentally: by adjusting the update through the inverse Hessian, we hypothesize that IHR gently rescales updates in sensitive directions rather than blocking them, while preserving those needed to learn new tasks.

\section{Conclusion}

We present Inverse Hessian Regularization (IHR), a novel memory-free method for continual learning in ASR. By correcting task-specific updates through a Kronecker-factored, layerwise inverse Hessian, IHR steers adaptation into directions less sensitive for previous tasks, thereby reducing forgetting while maintaining strong adaptability. As the inverse Hessian multiplication is applied only once per task, and only the estimate for the most recent task is retained, IHR remains lightweight in both computation and storage. Experiments on two CL benchmarks demonstrate consistent and significant improvements over state-of-the-art baselines, combining high adaptability with minimal forgetting. Ablation and analysis further illustrate the method’s effectiveness. Overall, these results highlight the promise of IHR as a principled and practical alternative to heuristic averaging for memory-free continual learning in ASR.

\vfill\pagebreak

% References should be produced using the bibtex program from suitable
% BiBTeX files (here: strings, refs, manuals). The IEEEbib.bst bibliography
% style file from IEEE produces unsorted bibliography list.
% -------------------------------------------------------------------------
\bibliographystyle{IEEEbib}
\bibliography{main}

\end{document}